\documentclass[10pt, conference]{IEEEtran}
\IEEEoverridecommandlockouts
\usepackage{cite}
\usepackage{amsmath,amssymb,amsfonts}
\usepackage{graphicx}
\usepackage{textcomp}
\usepackage{xcolor}
\usepackage{array}
\usepackage{tikz}
\usetikzlibrary{shapes.geometric, arrows}
\usepackage{algorithm}
\usepackage{algpseudocode}

\usepackage{float}
\usepackage{caption}

\tikzstyle{startstop} = [rectangle, rounded corners, minimum width=1.8cm, minimum height=0.8cm, text centered, draw=black, fill=red!30]
\tikzstyle{process} = [rectangle, minimum width=1.8cm, minimum height=0.8cm, text centered, draw=black, fill=orange!30]
\tikzstyle{decision} = [diamond, aspect=2, minimum width=1.8cm, minimum height=0.8cm, text centered, draw=black, fill=green!30]
\tikzstyle{arrow} = [thick,->,>=stealth]
\def\BibTeX{{\rm B\kern-.05em{\sc i\kern-.025em b}\kern-.08em
    T\kern-.1667em\lower.7ex\hbox{E}\kern-.125emX}}

\begin{document}

\title{MTDNS: Moving Target Defense for Resilient DNS Infrastructure}

\author{Abdullah Aydeger$^{\diamond}$, Pei Zhou$^{\diamond}$, Sanzida Hoque$^{\diamond}$, Marco Carvalho$^{\diamond}$ and Engin~Zeydan$^{\ast}$\\
$^{\diamond}$ Florida Institute of Technology, Melbourne, FL, USA, 32901. \\
$^{\ast}$Centre Tecnològic de Telecomunicacions de Catalunya (CTTC), Barcelona, Spain, 08860.\\
\protect Email: \{aaydeger, mcarvalho\}@fit.edu, \{pzhou2022, shoque2023\}@my.fit.edu, \\  engin.zeydan@cttc.cat 
}

\maketitle
\begin{abstract}

One of the most critical components of the Internet that an attacker could exploit is the DNS (Domain Name System) protocol and infrastructure. Researchers have been constantly developing methods to detect and defend against the attacks against DNS, specifically DNS flooding attacks. However, most solutions discard packets for defensive approaches, which can cause legitimate packets to be dropped, making them highly dependable on detection strategies. In this paper, we propose MTDNS, a resilient MTD-based approach that employs Moving Target Defense techniques through Software Defined Networking (SDN) switches to redirect traffic to alternate DNS servers that are dynamically created and run under the Network Function Virtualization (NFV) framework. The proposed approach is implemented in a testbed environment by running our DNS servers as separate Virtual Network Functions, NFV Manager, SDN switches, and an SDN Controller. The experimental result shows that the MTDNS approach achieves a much higher success rate in resolving DNS queries and significantly reduces average latency even if there is a DNS flooding attack.   
\end{abstract}



\section{Introduction}

In the constantly growing interconnected world of cyberspace, the Domain Name System (DNS) serves a critical part by enabling the conversion of human-readable domain names (e.g., www.example.com) into machine-readable IP addresses (e.g., 192.168.0.1). DNS has been used for various reasons and DNS servers can be placed in remote servers as well as edge servers \cite{harrilal2023bringing, hanna2023performance}. The significance and visibility of DNS make it a primary target for malicious cyber-attacks, specifically Distributed Denial of Service (DDoS) attacks. In a DDoS attack, by deliberately disrupting DNS servers, attackers impair the machines' capacity to establish a connection with a website, consequently rendering websites inaccessible to users. This disruption results in revenue losses, damaged reputations, and user dissatisfaction, which aligns with the attackers' goal of causing chaos for their targets. Moreover, a significant concern regarding IoT devices is that they are vulnerable to being transformed into botnets, which can be utilized for launching DDoS attacks~\cite{hayat2022ml}. This vulnerability originates from inherent weaknesses, such as the prevalence of weak or default passwords and inadequate security configurations, which grant unauthorized access to management systems and administrative controls. Based on Cloudflare's third DDoS threat report of 2023~\cite{Yoachimik_2023}, DNS-based DDoS attacks emerged as the predominant form of attack, constituting over 47\% of the total attacks. This signifies a 44\% surge in comparison to the preceding quarter of the year. 

Researchers are constantly involved in the pursuit of new methods and strategies to address and minimize the impact of DDoS attacks effectively. In recent years, researchers have shown keen interest in Software-Defined Networking (SDN)~\cite{bawany2017ddos} and Network Functions Virtualization (NFV)~\cite{yi2018comprehensive, aydeger2020cloud} based solutions for addressing DDoS attacks in DNS. SDN functionalities, including traffic analysis, dynamic packet forwarding rule configuration, and global network views, are widely utilized and leveraged to mitigate the challenges posed by DDoS attacks. At the same time, the NFV facilitates resource allocation and on-demand function instantiation flexibility and introduces new possibilities for mitigating DDoS attacks~\cite{zhou2017applying}. In addition, the Moving Target Defense (MTD)~\cite{aydeger2016mitigating, aydeger2018utilizing} techniques have emerged as a promising approach in the field of network security~\cite{huang2021trend}. The idea of MTD involves the dynamic alteration of network configuration and behavior \cite{saputro2020review}. Its purpose is to enhance uncertainty and seeming complexity for potential attackers, consequently decreasing their chances of success and raising the costs associated with their probing and attack activities \cite{aydeger2019moving}. 

\let\thefootnote\relax\footnotetext{This paper has been accepted for publication at  IEEE CCNC 2025. The copyright is with
IEEE and the final version will be published by IEEE.}
However, current approaches primarily combat DDoS attacks against DNS protocol and infrastructure by discarding or blocking suspicious traffic~\cite{xing2016defense, abou2020bringing, sagatov2023countering, gupta2018mitigating}. Certain solutions prioritize rerouting the traffic with network orchestration, which can result in a higher computational burden due to the added complexity of managing and coordinating multiple components, such as data centers and servers, within an orchestrated workflow~\cite{fayaz2015bohatei}. Additionally, these solutions may not be effective if the available server resources are exhausted. These works assume or try to understand and estimate the DNS clients' behavior as either malicious or legitimate \cite{nesary2022vdns}. Yet, how the defense mechanisms should work has not been investigated when we do not have clear and concise knowledge of the clients. Therefore, these methods rely on the detection of malicious traffic vs. legitimate, which carries the inherent risk of legitimate packet loss if the detection mechanism lacks accuracy.

In this paper, we propose \textit{MTDNS} that utilizes NFV with MTD techniques to mitigate DDoS attacks against the DNS servers. More specifically, the focus is on defending against flooding attacks by redirecting DNS traffic to a backup DNS server upon detecting a change in the network traffic pattern. The proposed approach offers increased flexibility for real-time and adaptable modifications while simultaneously reducing the computational burden and minimizing the risk of legitimate packet loss. The contributions of this paper can be summarized as follows: (i) A resilient DDoS Defense mechanism that utilizes MTD is designed for DNS in general SDN environments; (ii) The proposed approach is implemented as an SDN application; (iii) This system is evaluated in a real-world setup.

\section{Related Work}
\label{related_work}
In recent years, researchers have paid attention to SDN-based solutions for combating DDoS attacks against DNS, which aim to mitigate such attacks, predominantly encompassing flooding and amplification attacks \cite{nesary2022vdns}. The literature on this domain varies, with some studies concentrating on detection, some on mitigation, and a few addressing both. 

\subsection{Detection Strategies}
Detection strategies entail monitoring patterns in traffic to identify attacks. The threat model of the paper~\cite{ahmed2017mitigating} includes both flooding and amplification attacks and implements a Dirichlet process mixture model clustering algorithm to detect attack flows based on network statics in OpenFlow switches collected by the controller periodically. Xing et al.~\cite{xing2016defense} present a solution that detects DNS  attacks using packet rate and entropy calculation. While minimal resources are evident for this solution as it refrains from employing machine learning, the precision of detection can be lower with a constant threshold. On the other hand, Lyu et al.~\cite{lyu2021hierarchical} address the issue of DNS attacks on enterprise networks, utilizing trained and anomaly-based detection models to identify flooding attacks and scans from external sources of attacks instead of victims. They introduce a hierarchical and dynamic graph data structure for scalable monitoring to find the key attribute of the attacker. The WisdomSDN~\cite{abou2020bringing} proposes the technique derived from the MTD, involves mapping between DNS request and response, and utilizes the Bayes classifier to identify attacks based on entropy analysis.

\subsection{Mitigation Strategies}
The network's response to the detected threat to reduce the attack's impact can be referred to as a mitigation strategy. Mitigation strategies aim to provide effective countermeasures to minimize downtime and preserve service availability. WisdomSDN~\cite{abou2020bringing} implements a DNS mitigation scheme that uses rate-limiting techniques for illegitimate DNS traffic and aims to mitigate attacks by dropping DNS requests systematically if the packet rate of illegitimate DNS requests surpasses the threshold. The defense mechanism proposed by Xing et al.~\cite{xing2016defense} aims to mitigate the attack by dropping packets likewise. 
Sagatov et al.~\cite{sagatov2023countering} propose a mitigation scheme to counter these attacks that employs a temporary traffic-blocking technique from the attack's intermediary server as a mitigation strategy and also suggests another mechanism that involves executing a series of statistical rules for the number of detected packets. Gupta et al. propose a solution~\cite{gupta2018mitigating} that uses geographically distributed SDN routers and filters incoming traffic, dropping attack packets and allowing only legitimate traffic to reach its destination to minimize the impact of DNS flooding attacks.   
However, the mitigation scheme's reliance on dropping packets carries the inherent risk of also dropping legitimate packets and can potentially impact legitimate users by slowing down or blocking their DNS requests. 
Bohatei~\cite{fayaz2015bohatei} represents an initial research endeavor that employs the usage of SDN and NFV to mitigate DDoS attacks. The primary emphasis of this work is placed on network orchestration to defend against attacks rather than dropping packets. This approach can lead to higher computation overhead due to the additional complexity involved in managing and coordinating multiple components, such as data center and server, in an orchestrated workflow. 
In this study, we implement an approach that employs DNS server migration by considering MTD methods as a response to a potential attack. This approach is not only more flexible for immediate and dynamic adjustment, but it also requires less computation overhead and can avoid the loss of legitimate packets.

\begin{figure*}[]
\centering
\includegraphics[width=.62\linewidth]{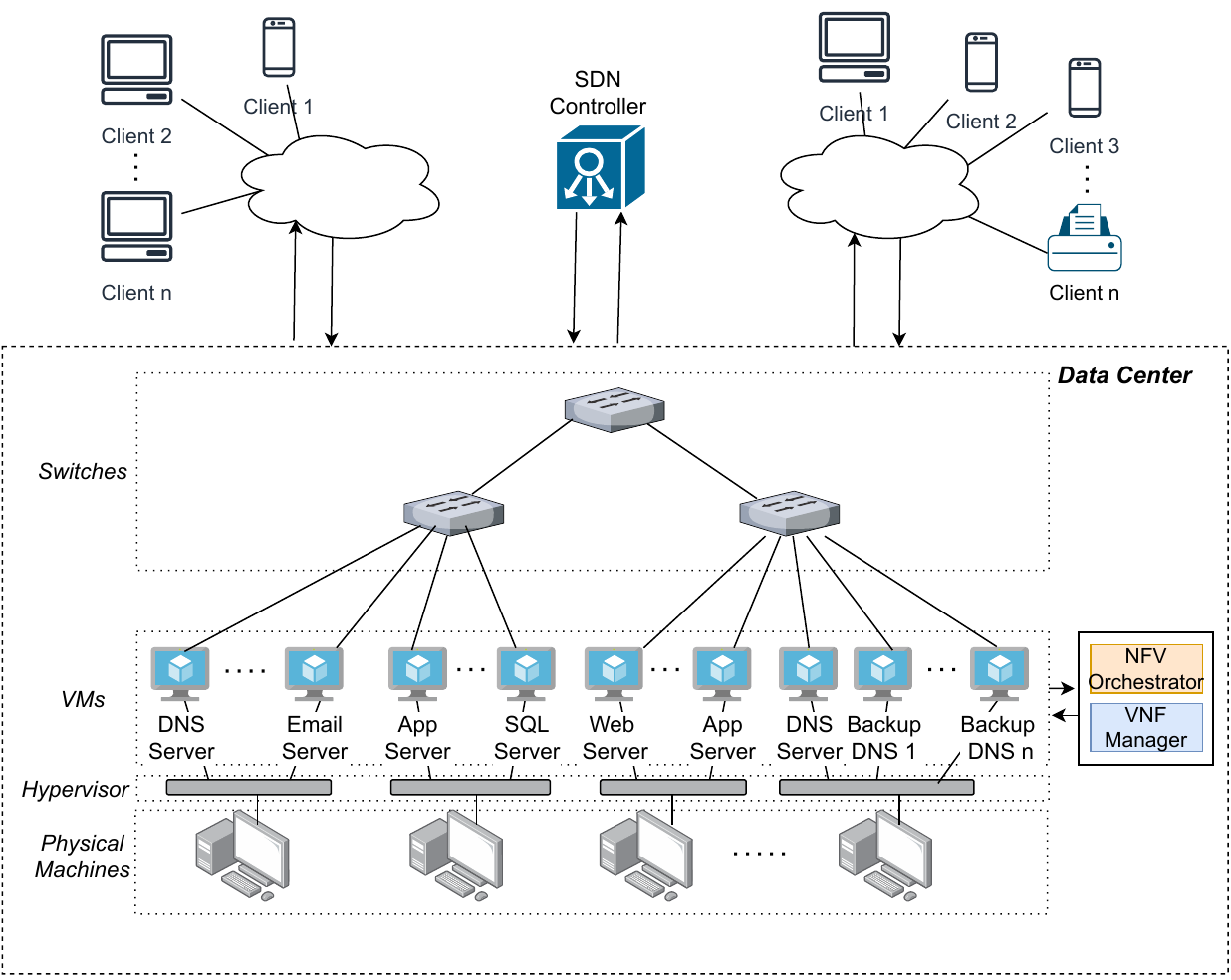}
\caption{System Model} 
\vspace{-0.1in}
\label{fig:model}
\end{figure*}

\section{Proposed Approach}
\label{solution}
In this section, we first describe the system model. Then, we explain the proposed approaches in the system model and provide the details of the algorithms we design and implement as our MTD-based defense mechanisms.


\subsection{System Model}
In the system model, we design and develop a resilient and adaptable strategy to mitigate the impact of DDoS attacks on DNS servers (i.e., DNS flooding). The key components of the system are depicted in Fig. \ref{fig:model} and explained as follows:

\begin{itemize}
    \item \textbf{Default DNS Server:} This is the current DNS server responsible for routine DNS resolution and service provisioning under normal traffic loads.
    \item \textbf{Backup DNS Server:} This server serves as an additional DNS server provisioned as a Virtual Network Function (VNF). It gets deployed automatically in response to triggered increasing DNS traffic to reduce the burden on the default DNS server. 
    \item \textbf{Controller:} The SDN controller is responsible for managing the traffic among the DNS servers as VNFs within the network. 
     \item \textbf{VNF Manager:} The VNF manager automatically deploys a new DNS server as a VNF when traffic exceeds a predefined threshold, triggered by controller notifications. As a part of the NFV orchestrator, the VNF manager is responsible for managing NFV services and integrating new network services as well as VNF packages, which enhance the interoperability of SDN components.
    
\end{itemize}

\subsection{MTD on DNS for Attack Mitigation}

\begin{figure}[]
\centering
\includegraphics[width=.9\linewidth]{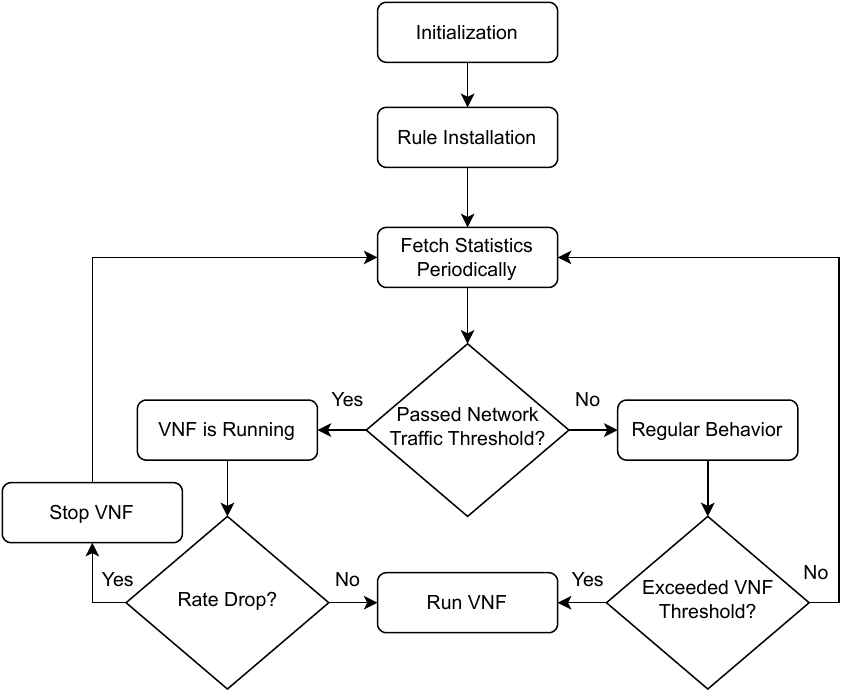}
\caption{Controller Flow} 
\vspace{-0.2in}
\label{fig:Flow}
\end{figure}

We design an MTD strategy to mitigate the consequences and risks of DNS flooding attacks. This process consists of three parts: matching flows, setting new routing rules and actions, and triggering the VNF manager. The workflow, encompassing these parts, is illustrated in the Figure \ref{fig:Flow} and proceeds as follows: 

\begin{enumerate}
    \item First, the system initializes the controller, and the controller sets a rule to direct DNS traffic to the default DNS server. 
    \item Second, the controller periodically fetches the packet count statistics from the previously set DNS request rule applied to the OVS. We employ the OpenFlow Group rule to implement the mitigation scheme. We use the packet count field in the request rule as the indicator of the occurrence of flooding attacks and the requirement for a VNF manager trigger. For this reason, the controller calculates the transmission rate of the DNS packets by dividing the packet difference by the time window of fetching statistics. The value of the transmission rate is used as the threshold for detecting the occurrence of suspicious DNS traffic. 
    \item Third, the controller triggers the VNF Manager, which then generates the backup DNS server. At this step, the controller installs new rules that handle the DNS request and response messages forwarded to/generated by the backup DNS server. 
    \item Fourth, if the controller detects a decrease in the packet rate compared to the previous packet rate above a predetermined threshold for rate drop (set at 40\% in this study), it triggers the VNF Manager. The allocated resources are deallocated, and OVS switches are updated to route to the default DNS server again.
\end{enumerate}

\subsection{DNS as VNF}
The procedures of the DNS Server as VNF is designed as follows:
\begin{enumerate}
    \item The controller triggers the VNF Manager, which then initializes a new backup DNS server once the network traffic threshold is passed.
    \item Whenever the critical threshold is met, the backup server is ready to handle the DNS traffic. Thus, the network traffic is not exposed to any additional delay.
\end{enumerate}
The startup and shutdown process of the DNS automation is presented in Algorithm \ref{alg:VMstartup} and Algorithm \ref{alg:VMshutdown}, respectively.



\begin{algorithm}[h]
\caption{DNS Startup Procedure}
\label{alg:VMstartup}
\begin{algorithmic}[1]
    \State Start processing event data
    \For{stat in event}
        \If{stat.PacketCount}
            \State Calculate rate
            \If {\(rate > T1\) \&\& DNS is inactive}
                \State Start DNS
                \State DNS state check
            \If{\(rate > T2\)}
                \State Initiate mitigation procedures
               \EndIf
            \EndIf
        \EndIf
    \EndFor
\end{algorithmic}
\end{algorithm}
\vspace{-0.12in}

\begin{algorithm}
\caption{DNS Shutdown Procedure}
\label{alg:VMshutdown}
\begin{algorithmic}[1]
    \For{group statistic}
        \If{group ID}
            \State Retrieve packet counts for buckets
            \State Calculate rates for each bucket
            \If{rate dropped}
                \State Terminate mitigation
                \State Switch to default
                \State Shutdown backup DNS
            \EndIf
        \EndIf
    \EndFor
\end{algorithmic}
\end{algorithm}
\vspace{-0.15in}
\subsection{Zone File Migration}
We also consider performing a zone file migration to maintain consistency in the DNS zone. The DNS zone file is essential as it stores the authoritative name server of a domain. If the backup DNS server does not have the same zone file as the default DNS server, the custom records that are configured in the default DNS server will not be recognized by the backup DNS server and vice-versa. Thus, we also do the zone file transfer while the new DNS is run.


\section{Evaluation}
\label{ex}
In this section, we introduce the experimental environment and setup that we use to implement our proposed methodology. These experiments are conducted using traditional DNS protocol. 
The experimental metrics and the results are provided subsequently. The source code and scripts developed for the realization of this project are available on Github \footnote{Gradresearch: Dns mtd,” Online, Dec 2023. [Online]. Available: https://github.com/Water-Meloon/GradResearch}.

\subsection{Setup Overview}
The hardware, operating system, OVS, and SDN, as well as the DNS server and structure details used in our experimental setup, are described in this subsection.

\textbf{Hardware and Operating System}: We have Dell desktops with an i7 CPU with 20 cores, 32GB RAM, and 512GB SSD storage. For the SDN setup, we configure the desktop as the OVS switch. 
Ubuntu server version 20.04.6 LTE is used as our operating system.



\textbf{OVS and SDN controller}: We use Open vSwitch version 2.13.8 and RYU as SDN controller for the experiment. The MTD logic is written directly in RYU. The controller's job is to continuously monitor DNS flows and write rules to respond to suspicious or simply increasing DNS requests and provide resilience. We write our logic directly for the controller instead of retrieving statistics from the API.


\textbf{Host VMs:} We emulate the situation where a legitimate user sends DNS requests to the default DNS server, and an attacker overwhelms the default DNS server. We have created two VMs for this situation. VM Host 1 is for the client, and VM Host 2 is for the attacker. We use Hping3~\footnote{http://wiki.hping.org/} and DNSperf~\footnote{https://www.dnsperf.com/} on the attacker and client machines. The DNS packets sent by clients using DNSperf have an average packet size of 28 bytes for requests and 70 bytes for responses. We also use a fixed client count of 125 and a duration of 20 seconds. For Hping3, we use randomly generated IP addresses to perform the DNS flooding attack. We set a payload of 120 bytes and various queries per second (QPS) for the DNS requests via Hping3. We obtain average values for each result by running ten tests with the same parameters and taking the average. 

\textbf{DNS Server VM:} We use the structure of BIND9~\cite{jinmei2006implementation} to implement our DNS servers as VNFs using VMs and VNF Manager. We set DNS servers as recursive resolvers. We use Virtualbox as a VM manager as it can connect VMs to virtual taps created on a virtual bridge. Since the DNS servers are VNFs, we install the two DNS servers (i.e., DNS resolvers) on two VMs managed by VNF Manager (i.e., Virtualbox). To do this, we create two virtual Ethernet adapters on our host machine, and two virtual DNS servers are connected to the OVS bridge via these two virtual taps. One of the DNS servers acts as a backup DNS resolver when needed. 

\textbf{Backup DNS VM:} As described in the previous section, the backup DNS VM helps the default DNS server unload DNS traffic when needed. We have two preset thresholds. One is used to trigger the start of the backup DNS server. Another is for forwarding traffic to the newly generated VNF (i.e., the backup DNS). The first threshold is calculated by taking half of the load balancing threshold. We start the backup server first to avoid the startup latency caused by the VNF boot process. The flow of the backup DNS VM is that the backup server is started when the VM startup threshold is exceeded. When the critical traffic load threshold is exceeded, the DNS traffic between the two shared servers is established by an OpenFlow group rule.

\textbf{VM Automation Using Vbox}:
We implement automation using the VBox command line manager called "VBoxManage." It allows us to start and save the state of a VM, making it an excellent choice for implementing automation. We emulate a middleware environment that controls the starting and stopping of the VM.

\textbf{SDN and OpenFlow}: 
We use SDN capabilities to implement our MTD approaches for DNS defense. OpenFlow rules have two fields that can be used to implement MTD in DNS. These are a field called Match and a field called Actions. The Match field is used to match flows with the defined parameters. In the case of MTD within DNS, we match the flows that use the UDP protocol, have port number 53, and are sent from the host VM to one of the DNS servers. As the name suggests, the action field defines actions when a data flow is matched with defined parameters. In MTDNS, we set the actions with which the responses from the backup DNS servers are to be handled when the mitigation procedure is applied.

\textbf{Zone File Migration}: We use the master-slave DNS configuration (i.e., primary-secondary DNS) to migrate the zone file. The idea is that the master server can transfer its zone file to its slave servers. The slave servers cannot modify the zone file received from the master server, which is read-only. The zone file transfer can be implemented by configuring a DNS server as the master server and the "allow transfer" option in the DNS server configuration file. This option allows the admin to specify which IP address or IP range the zone file is going to. The admin may also set the "also-notify" option to notify the slave servers that the zone file is reloaded. This option requires the slave servers to set the "allow-notify" option to accept the notification from the master server properly.

\subsection{Evaluation Metrics}
The following metrics are used to evaluate the performance of the proposed approach:
\begin{enumerate}
    \item \textbf{Average Latency:}
     The average time required for a DNS server to respond to a host's requests. This metric measures the DNS resolution with and without the SDN Controller (i.e., Ryu running) when a flooding attack occurs. 
     
    \item \textbf{Query Completion Rate:}
    This metric measures the proportion of DNS queries that a DNS server successfully resolves out of the total number of queries it receives within a certain period of time.
    This metric is obtained by using DNSperf.
    
    \item \textbf{Attack Mitigation Time:} This metric measures the time taken by MTDNS to respond and mitigate a sudden spike in incoming requests indicative of a potential attack. The data is derived by plotting the rate of network traffic in DNS servers while MTDNS is in operation, which demonstrates the time when the new VNF (i.e., backup DNS server) is initialized, and the network traffic is shared between the default and backup server. 
\end{enumerate}

\subsection{Results}
\label{result}

This section analyses the results and compares the performance of the DNS server with and without the proposed MTDNS approach. Table \ref{table:2} shows the results for DNSperf of DNS Server without the proposed MTD approach. The table shows that average latency increases, and the query completion rate drops as the number of clients increases. This is because increasing the number of clients raises traffic, which may result in delays owing to congestion and additional queuing and processing time. According to the table, without MTDNS, we obtain 92\% as the highest query completion rate. The effect of a DNS flooding attack becomes more visible with higher QPS, resulting in a 77\% query completion rate and an average latency of 4.579ms.

\begin{table}[ht]
\caption{Average Latency and Query Completion Rate of DNS servers without MTDNS}
\centering
\begin{tabular}{ |c|c|c|c| }
\hline
QPS  &  Query Completed & Avg Latency(ms) \\
\hline
50000  &  92\% & 3.665 \\ 
\hline
100000  &  91\% & 3.303\\
\hline
150000  &  77\% & 4.579 \\ 
\hline
\end{tabular}
\label{table:2}
\end{table}

Table \ref{table:3} shows the results for the MTDNS (i.e., the proposed approach) obtained with SDN and emulating a flooding attack with Hping3. The results show that our proposed method can bring the query fulfillment rate back to about 99\%. The average latency has seen a slight increase with the increasing QPS.

\begin{table}[ht]
\caption{Average Latency and Query Completion Rate of DNS servers with MTDNS}
\centering
\begin{tabular}{ |c|c|c|c| }
\hline
QPS &  Query Completed & Avg Latency(ms) \\
\hline
50000   & 99\% &0.966 \\ 
\hline
100000   & 99\% &1.006\\
\hline
150000 & 99\% & 1.005\\
\hline
\end{tabular}
\label{table:3}
\end{table}

\begin{figure}[htbp]
\centering
\includegraphics[width=\linewidth]{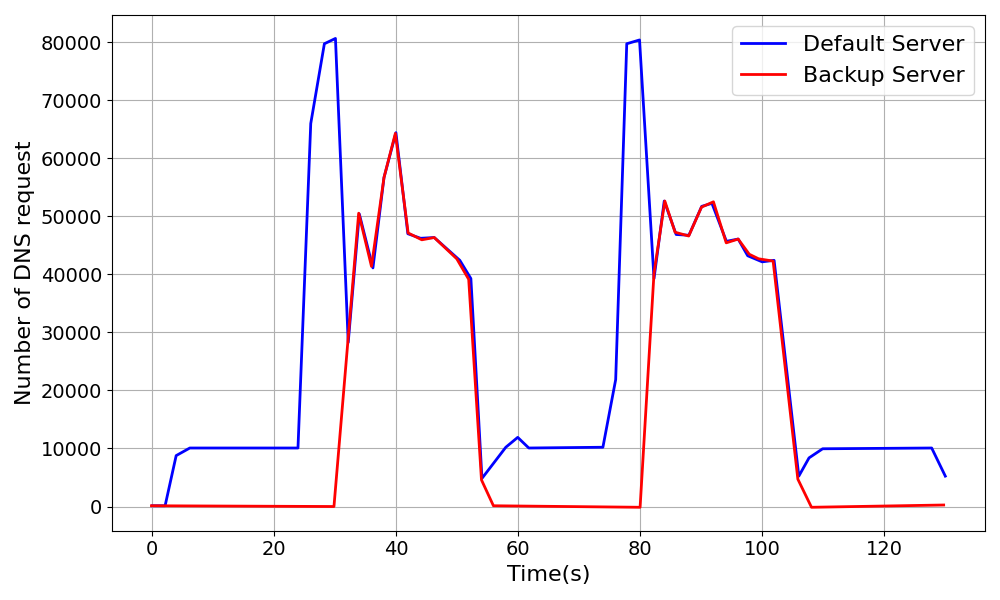}
\vspace{-0.2in}
\caption{Network Traffic in DNS Servers while MTDNS is running} 
\vspace{-0.1in}
\label{fig:PacketRate}
\end{figure}

Figure \ref{fig:PacketRate} visualizes alleviating the load on the DNS servers when the MTDNS is activated. We use the default packet rate calculated from the default DNS request rule before the execution of the MTDNS functions and the default DNS server bucket packet rate from the group rule. Before the first spike, the packet rate becomes stable, around 10000 QPS. The spike indicates that the packet rate exceeds the threshold to trigger VNF Manager. The drop after the first spike indicates the start of the MTDNS, which has mitigated the attack and provided resiliency. The red line represents the packet rate for the backup DNS server. It remains zero before the group rule is applied (i.e., VNF has not been needed). Once the group rule is installed, we fetch the backup DNS server bucket packet count for the MTD packet rate calculation. When the group rule and the group are deleted, the packet rate becomes zeros again. Those zeros represent that the backup DNS server is not needed, and the resources are deallocated. The two spikes on the graph are mitigated by using MTD. The initial MTDNS detection and activation take place at the 30-second mark, and the traffic is mitigated within around 2.1 seconds. The second mitigation is initiated at approximately the 80th second, resulting in mitigation in approximately 2.3 seconds. 

\section{Conclusion}
\label{conclusion}

In this paper, we propose and discuss the feasibility of combining DNS with NFV and SDN to provide resiliency against DNS flooding attacks. We illustrate how proposed MTD approaches work when the SDN controller detects a high packet rate at a specific time. We explain how the controller switches between the default DNS server and the backup DNS server to route the incoming DNS requests on the run time via VNF Manager. Finally, our experimental findings demonstrate the benefits and practicality of MTDNS. This integration result in a much higher success rate in resolving DNS queries, a notable decrease in average latency, and a visible mitigation of anomalous increased traffic. 

Our work considers only traditional DNS, which utilizes UDP port 53. In our future work, we plan to incorporate newer DNS protocols, such as DNS Security and DNS over Transport Layer Security (TLS).





\bibliographystyle{IEEEtran}
\bibliography{ref}
\end{document}